\documentclass[aps,prl,twocolumn]{revtex4-1}
\usepackage{graphicx,color}
\usepackage{dcolumn}
\usepackage{bm}
\usepackage{amsmath,amsthm,amssymb}
\usepackage{gt13}

\begin{document}

\title{
Thermal vector potential theory of transport induced by temperature gradient
}

\author{Gen Tatara}
\affiliation{%
 RIKEN Center for Emergent Matter Science (CEMS)\\  
2-1 Hirosawa, Wako, Saitama 351-0198, Japan
}

\begin{abstract}
A microscopic formalism to calculate thermal transport coefficients is presented based on a thermal vector potential, whose time-derivative is related to a thermal force.
The formalism is free from unphysical divergences reported to arise when Luttinger's formalism is applied naively, because the equilibrium (\textquoteleft diamagnetic\textquoteright) currents are treated consistently.
The mathematical structure for thermal transport coefficients are shown to be identical with the electric ones if the electric charge is replaced by energy. 
The results indicates that the thermal vector potential couples to energy current via the minimal coupling.
\end{abstract}

\date{\today}

\maketitle

\newcommand{\alphaG}{\alpha_{\rm s}}
\renewcommand{\betana}{\beta_{\rm s}}

\newcommand{\Jl}{J_0}
\newcommand{\Dl}{D_0}
\newcommand{\al}{a}
\newcommand{\boson}{b}
\newcommand{\Aad}{A_{\rm s}}
\newcommand{\Aadv}{\Av_{\rm s}}
\newcommand{\AU}{A_U}
\newcommand{\AUv}{\Av_U}
\newcommand{\epk}{\epsilon_{k}}
\newcommand{\gammak}{\gamma_k}
\renewcommand{\gyro}{\gamma_{\rm g}}
\newcommand{\HAT}{H_{A_T}}
\newcommand{\Hdwsw}{H_{\rm st}}
\newcommand{\intrv}{\int{d^3r}}
\newcommand{\inttd}{\int{d^2r}}
\newcommand{\jmag}{{j}_{\rm m}}
\newcommand{\jmagv}{\bm{j}_{\rm m}}
\newcommand{\jmagtd}{{j}_{\rm m}^{\rm (2d)}}
\newcommand{\jmagvtd}{\bm{j}_{\rm m}^{\rm (2d)}}
\newcommand{\kappatd}{\kappa^{\rm (2d)}}
\newcommand{\kappasw}{\kappa_{\rm sw}}
\newcommand{\Kperp}{K_\perp}
\newcommand{\sumrv}{\int\frac{d^3r}{a^3}}
\newcommand{\sumtd}{\int\frac{d^2r}{a^2}} 
\newcommand{\sumkom}{\sum_{\kv\omega}}
\newcommand{\setil}{\tilde{\se}}
\newcommand{\Simptil}{\tilde{S}^{\rm (i)}}
\newcommand{\ua}{u_{\rm c}}
\newcommand{\uT}{u_T}

\renewcommand{\Simpv}{{{\bm S}^{\rm (i)}}}
\renewcommand{\Simp}{{S^{\rm (i)}}}

\newcommand{\ATv}{\Av_{T}}

Conversion of heat into electric and other currents and vice versa is of essential importance from the viewpoint of realizing devices with low energy consumption.
Of recent particular interest is heat-induced spin transport in the field of spintronics, where 
spin current is expected to lead to novel mechanisms for information technology, and to devices with low-energy consumption due to the absence or weak Joule heating. 

A hot issue in spintronics is to use magnetic insulators, which are suitable for fast magnetization switching and low-loss signal transmission.
Insulators have, however, a clear disadvantage that electric current cannot be used for its manipulation. Instead, temperature gradients become the most important driving force in inducing spin transport. 
To study thermally-induced spin transport theoretically, a microscopic formulation is necessary for full understanding and for quantitative predictions. 
A microscopic description is, however, not straightforward; temperature gradients and thermal forces are macroscopic quantities arising after statistical averaging, and thus it is not obvious how to represent those effects in a microscopic quantum mechanical Hamiltonian.

In 1964, Luttinger proposed a solution \cite{Luttinger64}.
To describe the effect of temperature gradient, he introduced a scalar potential $\Psi$, which he called a \textquoteleft gravitational\textquoteright \ potential, which couples to energy density of the system, ${\cal E}$, via an interaction Hamiltonian, 
\begin{align}
  H_{\rm L} &= \intrv \Psi {\cal E}.
  \label{HTLuttinger}
\end{align}
Although the microscopic origin of the potential has not been addressed, he argued that 
to satisfy the Einstein relation the potential adjusts itself to balance the thermal force, resulting in an identity 
$\nabla \Psi= \frac{\nabla T}{T}$ in the thermal equilibrium. 
Owing to this trick, thermal transport coefficients can be calculated by linear response theory with respect to the field $\Psi$ without introducing the temperature gradient in a microscopic Hamiltonian.
Another approach, based on the Landauer-B\"uttiker formalism, was presented by Butcher \cite{Butcher90}. 

Luttinger's method has been applied to study various thermally-induced electron transports \cite{Smrcka77,Oji85,Catelani05,Michaeli09,Qin11,Schwiete14,Eich14}, magnon transport \cite{Matsumoto11a} and thermally-induced torque \cite{Kohno14}.
It turned out, however, that naive application often leads to apparently wrong transport coefficients which diverge as $T\ra0$ \cite{Smrcka77,Oji85,Qin11}.
In the case of the thermal Hall effect, the divergence was identified to be due to a wrong treatment of the equilibrium diamagnetic current induced by the applied magnetic field, and it was found that the physical Hall coefficient is obtained if one subtracts the equilibrium magnetization current before applying linear response theory \cite{Qin11}. 
A similar problem was reported recently for thermally-driven spin-transfer torques \cite{Kohno14}. 

In the case of electrically-driven transport, elimination of unphysical equilibrium contribution from transport coefficients is guaranteed by U(1) gauge invariance, which represents charge conservation. 
In the presence of an electromagnetic vector potential, $\Av$, the physical electric current  has two components, a paramagnetic current (the first term) and a diamagnetic current (the second term), as 
$\jv=\frac{e}{m}\average{\hat\pv}-\frac{e^2}{m} \nel \Av$, where $\average{\hat\pv}$ is quantum average of the momentum density operator, $\nel$ is electron density and $e$ and $m$ are electron's charge and mass. 
The paramagnetic current contains an equilibrium contribution arising from all the electrons below the Fermi level, which turns out to be $\frac{e^2}{m} \nel \Av$. This equilibrium contribution thus cancels perfectly with the diamagnetic contribution, leaving only the contribution from excitations in the transport coefficients \cite{Rammer86}. 
Obviously, a consistent treatment of the two contributions is necessary for the cancellation of equilibrium contribution and for gauge-invariant physical results. 
If one uses, instead of a vector potential, a scalar potential to describe an conservative electric field, the role of the diamagnetic current is not clearly seen, and wrong results easily arise if an inconsistent treatment is employed.

From those experiences in electrically-induced transport, the divergence in the thermally-induced transport  described by Luttinger's $\Psi$ is expected to be due to an incorrect treatment of the \textquoteleft diamagnetic\textquoteright\  contribution.\cite{GT15diamag}
What Qin et al. \cite{Qin11} showed is that the \textquoteleft diamagnetic\textquoteright\ effect is consistently taken account of if divergenceless magnetization current is included. 
If one could construct a vector potential representation of thermal effects, the \textquoteleft diamagnetic\textquoteright\ effect would be treated consistently and straightforwardly, since the \textquoteleft diamagnetic\textquoteright\  current is defined by the Hamiltonian and the vector potential without ambiguity.  

Temperature gradients exert a  statistical force proportional to $\nabla T$, which is conservative, i.e., has no rotation component.
Still, one may introduce a rotationless vector potential to describe the force. 
In the  case of classical charged particles described by a Hamiltonian 
$H=\frac{(\pv-e\Av)^2}{2m}+V(\rv)$ ($V$ is a scalar potential), the total force is $\Fv=-e\frac{\partial \Av}{\partial t}-\nabla V$ if $\nabla \times \Av=0$.
Any rotationless force can thus be represented by use of a vector potential without introducing a scalar potential.
In thermally-driven transport, a thermal force proportional to $\nabla T$ is represented by a vector potential which we call the thermal vector potential. 

The objective of this paper is to propose a formalism describing thermal effects by a thermal vector potential, and to demonstrate that the formalism works perfectly for a few simple cases of thermally-driven electron and energy transport, without yielding unphysical divergences.
Since the \textquoteleft charge\textquoteright\  to which the temperature gradient couples is energy, the thermal vector potential couples to the energy current density operator, $\jv_{\cal E}$.
We first carry out a derivation of a thermal vector potential form of the interaction Hamiltonian by looking for a Hamiltonian equivalent to the Luttinger's Hamiltonian.
Local thermal equilibrium is assumed. 
We then derive expressions for electric current and energy current by use of conservation laws, and identify the \textquoteleft diamagnetic\textquoteright\  currents. 
It is shown that the \textquoteleft diamagnetic\textquoteright\  currents remove unphysical equilibrium contribution to transport coefficients. The results satisfy the Wiedemann-Franz law.
We shall demonstrate that the obtained expressions for the currents indicates the minimal coupling of the thermal vector potential.

It was recently demonstrated by Shitade that a gauge theory of gravity constructed imposing local space-time translation symmetry contains a vector \textquoteleft gauge field\textquoteright\ as well as a scalar potential corresponding to that of Luttinger \cite{Shitade14}.
The model was applied to describe thermal transport of non-interacting electrons and Wiedemann-Franz law was shown to be satisfied. 
The origin of local translation symmetry in the context of thermal transport was not addressed to. 

In this paper, we use energy conservation law  to derive a vector potential representation of thermal effects, without assuming invariance under local space-time translation. 
We start with rewriting the Luttinger's Hamiltonian  
by use of continuity equation for operators ${\cal E} $ and $\jv_{\cal E}$, 
\begin{align}
  \dot{\cal E} &= -\nabla\cdot \jv_{\cal E}.
  \label{energyconsev}
\end{align}
 as 
\begin{align}
  H_{\rm L} (t) 
  &= \intrv \int_{-\infty}^t dt'\jv_{\cal E}(t')\cdot \nabla\Psi(\rv,t) 
  \label{HTLint}
\end{align}
where we used Gauss's theorem assuming that no field exists at $r\ra\infty$.
This expression is not of the form of an interaction between a vector potential and energy current because of the time-integration.
We here look for a Hamiltonian $\HAT$ which agrees with Eq. (\ref{HTLint}) when long time average is considered, namely, $\int _{-\infty}^\infty dt \HAT = \int _{-\infty}^\infty dt H_{\rm L} (t) $.
The result is   
\begin{align}
 \HAT
  &\equiv -\intrv \jv_{\cal E}(\rv,t)\cdot \Av_{T}(t)
  \label{HATdef}
\end{align}
where $ \Av_{T}(t)\equiv \int^t_{-\infty} dt' \nabla\Psi(t')$ \cite{GT15AT}
is the thermal vector potential, which satisfies 
\begin{align}
   \partial_t \Av_{T}(\rv,t) 
    = \nabla\Psi(\rv,t)=\frac{\nabla T}{T}.
\end{align}

The interaction Hamiltonian (\ref{HATdef}) is understood as representing the thermodynamic potential change when a static temperature gradient is applied.
In fact, the rate of the change of the entropy (${\cal S}$) due to an energy current is \cite{Landau84}
\begin{align}
  \dot{\cal S} &=-\intrv \frac{1}{T}\nabla\cdot \jv_{{\cal E}} 
  =-\intrv\jv_{{\cal E}}\cdot\frac{\nabla T}{T^2} ,
\end{align}
and this entropy change modifies the thermodynamic potential, $E-T{\cal S}-\mu N$ ($E$ is the internal energy and $N$ is the electron number).
The effective Hamiltonian describing a DC thermal force, $H_{\cal S}\equiv -T{\cal S}$, is therefore (to the linear order in $\nabla T$)
\begin{align}
  H_{{\cal S}}
   =
   \frac{1}{T}\intrv \int_{-\infty}^t \jv_{{\cal E}}(t')dt' \cdot \nabla T .
   \label{HdeltaS}
\end{align}
This reduces to Eq. (\ref{HTLint}) after the replacement $\nabla\Psi\ra \frac{\nabla T}{T}$.

We now apply the thermal vector potential interaction, Eq. (\ref{HATdef}), to study thermal transport and demonstrate that the formalism works perfectly.
We consider free electrons with a quadratic dispersion, described by the Hamiltonian 
$
H_0  \equiv \intr  {\cal E}_0 $, 
where  
$
  {\cal E}_0  \equiv \frac{\hbar^2}{2m} (\nabla c^\dagger)(\nabla c)-\mu \cdag c
$
is the free electron energy density, $\mu$ is the chemical potential and $\cdag$ and $c$ are creation and annihilation operators of the electron, respectively. 
The energy current density is derived by use of the energy conservation law, Eq. (\ref{energyconsev}).
For free electrons, 
$\nabla\cdot \jv_{\cal E}^{(0)}\equiv - \frac{i}{\hbar}[H_0,{\cal E}_0(\rv)]$, 
and the result is \cite{GT15curl}
\begin{align}
\jv_{\cal E}^{(0)}
&= \frac{i\hbar^3}{(2m)^2}
\biggl[ (\nabla^2 c^\dagger) \nabla c - (\nabla c^\dagger)(\nabla^2 c)\biggr]
- \frac{\mu}{e} \jv^{(0)},
\label{jefree}
\end{align}
where $\jv^{(0)}\equiv \frac{-ie\hbar}{2m}c^\dagger \nablalr c $ is the paramagnetic part of the electric current density.
We focus on the uniform component of the current considering the case of spatially uniform temperature gradient, which reads  ($V$ is system volume)
\begin{align} 
  j_{{\cal E},i}^{(0)}
    &= \frac{\hbar}{m} \frac{1}{V} \sum_{\kv}  k_i \ekv c^\dagger_{\kv} c_{\kv} ,
 \label{jEfree}
\end{align}
where $\kv$ is a wave vector and 
$  \ekv  \equiv \frac{\hbar^2 k^2}{2m}-\mu$ is the energy measured from the Fermi energy.

We now apply this interaction to study thermally-driven longitudinal electron transport on a basis of diagrammatic (Green's function) formalism.
Besides the interaction Hamiltonian, we need to take account of the \textquoteleft diamagnetic\textquoteright\  current contribution proportional to $\Av_T$.
We derive it by use of the charge conservation law, $\dot{\rho}+\nabla\cdot\jv=0$ ($\rho$ is electric chage density) taking account of thermal vector potential. 
Namely, we calculate a commutator, $-\frac{ie}{\hbar}[\HAT^{0},\cdag c]\equiv \nabla\cdot\jv^{A_T}$, and derive the expression for $\jv^{A_T}$.
The result of the uniform component is
\begin{align} 
j_i^{A_T}  
 &= -\frac{e}{m} A_{T,j} \frac{1}{V}  \sum_{\kv}
  \gammak^{ij} 
  c^\dagger_{\kv} c_{\kv} 
 \label{jsfree2}
\end{align}
where 
\begin{align}
 \gammak^{ij} & \equiv \ekv\delta_{ij}+\frac{\hbar^2}{m} k_i k_j.
\end{align}

As expected from Eq. (\ref{jEfree}), the diagrammatic calculation is carried out by a straightforward replacement of charge $e$ in the electric field driven case \cite{Rammer86} by energy $\ekv$. 
Including the interaction with the vector potential to the linear order, the DC paramagnetic current is 
\begin{align}
j^{(0)}_i  
 &= \frac{e\hbar}{mV} \sumkom
 \epk^{ij} \biggl[
 \dot{A}_{T,j} \frac{f'(\omega)}{2}  (\phi_{\kv\omega})^2
       -2 {A}_{T,j}  f(\omega) \Im[(\ga_{\kv,\omega})^2]
    \biggr],
\end{align}
where $\epk^{ij} \equiv \frac{\hbar^2}{m}k_ik_j\ekv$, $ \dot{\Av}_{T}\equiv \frac{\partial \Av_{T}}{\partial t}$,  
$\phi_{\kv\omega}\equiv \ga_{\kv,\omega} - \gr_{\kv,\omega}$, 
$\sum_{\omega}\equiv \sumom $, $\Im$ denotes the imaginary part and 
$f(\omega) \equiv [e^{\beta\hbar\omega}+1]^{-1}$ is the Fermi distribution function 
($\beta\equiv(\kv T)^{-1}$, $\kb$ being the Boltzmann constant).
The retarded and advanced Green's functions for free electron are denoted by 
$
  \gr_{\kv,\omega} = \frac{1}{\hbar\omega-\ekv+\frac{i\hbar}{2\tau}}
$, 
and $\ga_{\kv,\omega}=( \gr_{\kv,\omega})^*$, 
where $\tau$ is the elastic lifetime.
By use of $\frac{\hbar k_j}{m}(\ga_{\kv\omega})^2 = \partial_{k_j}\ga_{\kv\omega}$ and integration by parts with respect to $\kv$, we rewrite the last contribution using  
\begin{align}
\sum_{\kv}  \epk^{ij} \biggl[ (\ga_{\kv\omega})^2-(\gr_{\kv\omega})^2 \biggr]
&= -
\sum_{\kv} \gammak^{ij} \phi_{\kv\omega} ,
\end{align}
to obtain 
\begin{align}
j^{(0)}_i  
 &= - \frac{e\hbar}{mV}  \sumkom
  \dot{A}_{T,j}  \epk^{ij} \frac{f'(\omega)}{2} (\phi_{\kv\omega})^2 
  - j_i^{A_T}  
      \label{j0res1}
\end{align}
where 
$
j_i^{A_T}  
 = i\frac{e\hbar}{mV} A_{T,j} \sumkom  \gammak^{ij}
 f(\omega)  \phi_{\kv\omega} 
$ 
agrees with the diamagnetic current (Eq. (\ref{jsfree2})). 
The equilibrium (diamagnetic) contribution is therefore eliminated from the physical thermally-induced electric current, obtaining 
$j_i = \sigma_T E_{T,i}$,
where 
\begin{align}
\Ev_{T}\equiv - \frac{\partial \Av_{T}(\rv,t) }{\partial t} =-\frac{\nabla T}{T}
, \end{align}
is the thermal field, and 
$
\sigma_T 
  \equiv 
  \frac{e\hbar^5}{6m^2 V\tau^2} \sumkom 
 k^2\ekv  
  f'(\omega) |\gr_{\kv\omega}|^4 
$ (assuming rotational symmetry for $\kv$).
The low temperature behavior is seen by a series expansion 
($\Phi_k(\omega)\equiv  |\gr_{\kv\omega}|^4$)
\begin{align}
 \int_{-\infty}^\infty d\omega f'(\omega)\Phi_k(\omega) 
 &= -\Phi_k(0) -\frac{\pi^2}{6}(\kb T)^2\Phi_k''(0), \label{Texpansion}
\end{align}
where $O(T^4)$ is neglected.
Since $\ekv=0$ on the Fermi surface, we have $\sumkv k^2\ekv \Phi_k(0)=0$ (to the leading order of $\frac{\hbar}{\ef\tau}$).
We therefore see that 
 $\sigma_T=O(T^2)$ at $T\ra0$ and 
the thermally-induced current vanishes at $T=0$.

We now study the thermally-driven Hall effect and show that the vector potential formulation does not lead to unphysical results like the one reported in the Luttinger's scheme \cite{Qin11}.
We introduce the interaction with an electromagnetic vector potential, $H_A\equiv -\intrv \Av\cdot\jv$, to describe the effect of the applied magnetic field, where 
$ \jv = \jv^{(0)}+ \jv^{A} +\jv^{A_T}$,  
$\jv^{A}\equiv -\frac{e^2}{m}\Av c^\dagger c$ being the diamagnetic current of the electromagnetic origin.
The electromagnetic vector potential is treated as static but has finite wave vector, since its role here is to represent a static magnetic field, $\Bv=\nabla\times \Av$.
The thermal vector potential has an infinitesimal angular frequency ($\Omega$) and is spatially uniform.
The Hall current is calculated to the lowest order, i.e., linear in both $\Av$ and $\Av_T$.
It turns out that the leading contribution is linear both in the angular frequency $\Omega$ and in the wave vector $\qv$.

\begin{figure}[tbh]
\includegraphics[height=3.5\baselineskip]{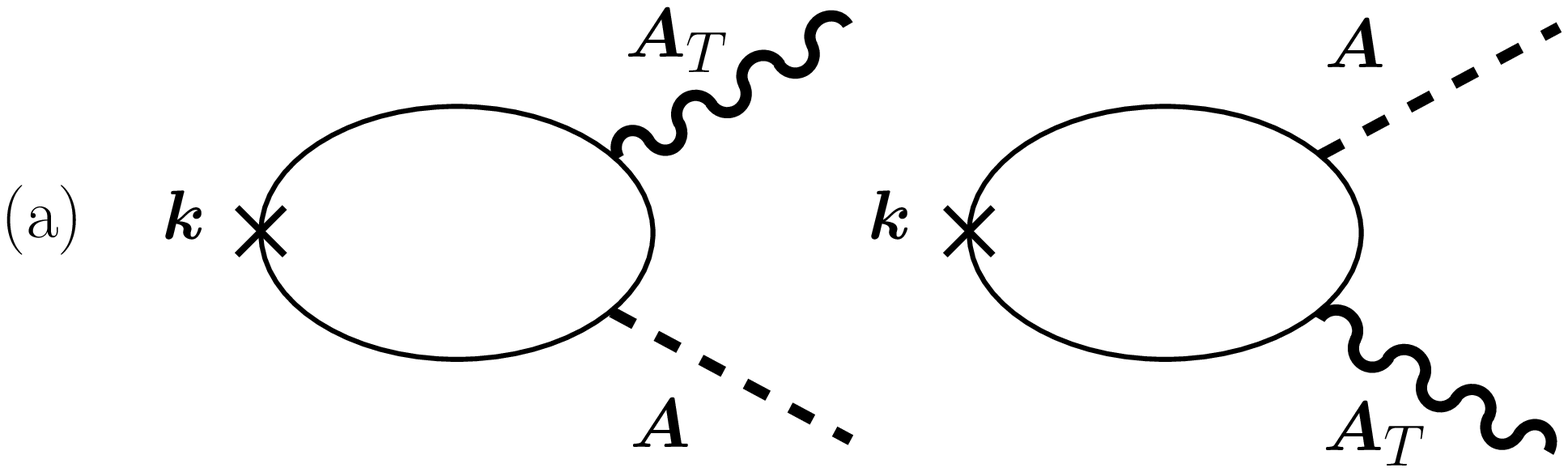}
\includegraphics[height=3.5\baselineskip]{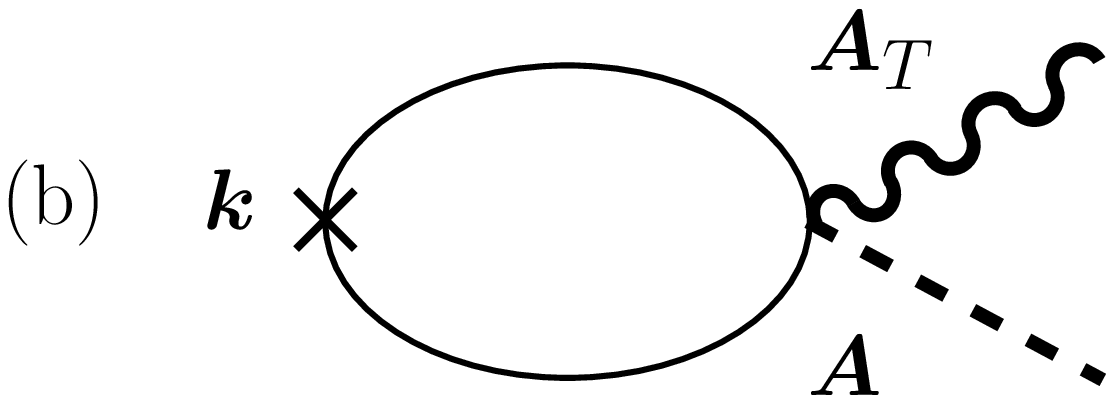}
\includegraphics[height=3.2\baselineskip]{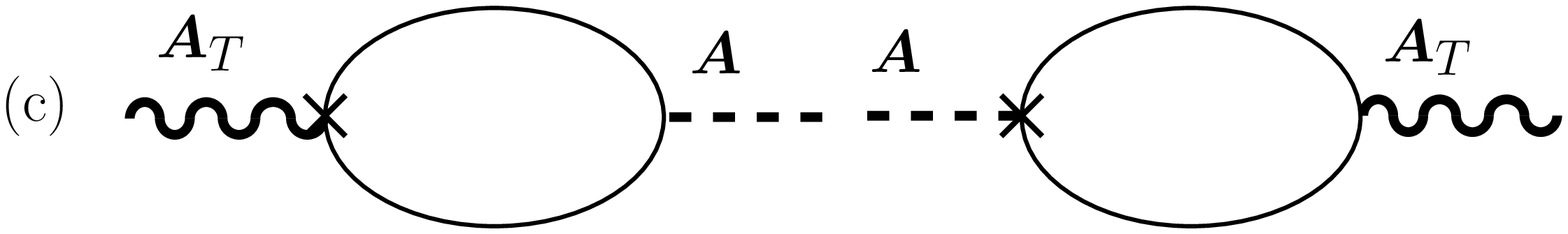}
  \caption{ Diagramatic representation of the contributions to the thermal Hall effect. 
  Solid, wavy and dotted lines denote the electron, thermal vector potential $\Av_T$ and the electromagnetic vector potential $\Av$, respectively.
  Diagrams (a) and (b) correspond to contributions of Eqs. (\ref{jhalla}) and (\ref{jhallb3}), respectively, while contributions of diagram (c) vanish.
  \label{FIGHall}}
\end{figure}

The contributions to the paramagnetic part of the current, $\jv^{(0)}$ are shown diagrammatically in Fig.  \ref{FIGHall}(a)(b). 
(\textquoteleft Diamagnetic\textquoteright\ currents shown in Fig. \ref{FIGHall}(c), vanish, since $\Av$ and $\Av_T$ carry only either finite angular frequency or finite wave vector in the present description.)
The contribution of Fig. \ref{FIGHall}(a)  is 
\begin{align}
j^{\rm (a)}_i  
 &= -\frac{e^2\hbar^4}{m^3 V} (\nabla_m A_j) \dot{A}_{T,l} \sumkom 
 k_i k_j \gammak^{lm} \Phi^{\rm (H)}_{\kv\omega},
 \label{jhalla}
\end{align}
where 
$ \Phi^{\rm (H)}_{\kv\omega} \equiv 
 \Im[f'(\omega)  (\gr_{\kv,\omega})^2 \ga_{\kv,\omega}
 + \hbar f(\omega) (\ga_{\kv,\omega})^4 ]
 $.

Because of \textquoteleft diamagnetic\textquoteright\  current due to the thermal vector potential, $\jv^{A_T}$, we have an interaction vertex, $-\intr \Av\cdot \jv^{A_T} $, linear in both $\Av$ and $\Av_T$.
The contribution shown in Fig. \ref{FIGHall}(b) 
arises from this interaction vertex. 
It is  
\begin{align}
j^{\rm (b)}_i  
 &= \frac{e^2\hbar^4}{m^3 V}  (\nabla_m A_j) \dot{A}_{T,l} \sumkom 
 k_i k_m  \gammak^{jl} \Phi^{\rm (H)}_{\kv\omega} .
 \label{jhallb3}
\end{align}
The total Hall current, $j^{\rm (H)} \equiv j^{\rm (a)}+j^{\rm (b)}$, is finally obtained as 
\begin{align}
\jv^{\rm (H)}_i  
  & = \Theta_H (\Ev_T \times \Bv),
 \label{jhall}
\end{align}
where $
\Theta_H \equiv 
 \frac{e^2\hbar^4}{3m^3 V} \sumkom 
 k^2 \epsilon_{\kv} \Phi^{\rm (H)}_{\kv\omega}
$.
We see that the Hall current vanishes at $T=0$ (see Eq. (\ref{Texpansion})), as is physically required.
The vector potential formalism applied straightforwardly therefore leads to the correct result, in sharp contrast to Luttinger's \textquoteleft gravitational\textquoteright\  potential formalism.

For consistency of the vector potential formalism, we need to confirm that \textquoteleft diagmagnetic\textquoteright\  current arises also for the energy current. 
This is not a trivial issue, since we cannot invoke a gauge invariance concerning the energy current, in contrast to the case of electric current.
In our scheme, \textquoteleft diamagnetic\textquoteright\  contribution is explored again by looking into the energy conservation law. 
In fact, including the thermal vector potential interaction, Eq. (\ref{HATdef}), in the left-hand side of Eq. (\ref{energyconsev}), we see that the energy current acquires a \textquoteleft diamagnetic\textquoteright\  contribution linear in $A_T$.
After a straightforward calculation, its uniform component is obtained as 
\begin{align}
  {j}_{{\cal E},i}^{A_T} 
    & = -\frac{1}{m}A_{T,j}\frac{1}{V}\sumkv \gamma_{T,k}^{ij}\cdag_\kv c_\kv ,
     \label{jEAT0}
\end{align}
where 
\begin{align}
  \gamma_{T,k}^{ij}
    & \equiv  \ekv \biggl(\ekv\delta_{ij}+\frac{2\hbar^2}{m}k_ik_j \biggr).
\end{align}
We see here that the matrix $\gamma_{T,k}^{ij}$ for this energy current correction satisfies 
$\gamma_{T,k}^{ij}= \frac{\partial}{\partial k_i} [k_j (\ekv)^2 ]$, and thus cancellation of the 
unphysical equilibrium contribution occurs, in the same manner as the electric currents discussed above.
The total energy current density induced by the thermal vector potential is 
$\jv_{{\cal E}}  = -\kappa \nabla T$, where 
$
 \kappa \equiv 
  \frac{\hbar}{2VT} \lt(\frac{\hbar}{m}\rt)^2  \sumkom  k_ik_j (\ekv)^2  f'(\omega) (\phi_{\kv\omega})^2 
$.
The coefficient satisfies the Wiedemann-Franz law, $\kappa= \frac{\pi^2}{3}\kb^2 T \sigmab$, where $\sigmab$ is the Boltzmann conductivity.

We have confirmed that thermal vector potential formalism applied for various thermally-induced transport phenomena leads straightforwardly to physical transport coefficients.
The uniform contributions to the electric and energy current densities we have derived are 
\begin{align}
 j_{i} 
    &= \frac{e\hbar}{m} \frac{1}{V} \sum_{\kv} 
   \biggl[  k_i  -eA_i-\gammak^{ij}A_{T,j} \biggr]
    c^\dagger_{\kv} c_{\kv} \nonumber \\
 j_{{\cal E},i} 
    &= \frac{\hbar}{m} \frac{1}{V} \sum_{\kv} 
   \biggl[  k_i \ekv -e\gammak^{ij}A_j-\gamma_{T,k}^{ij}A_{T,j} \biggr]
    c^\dagger_{\kv} c_{\kv} . \label{jEresult}
\end{align}
The key for direct access to physical results in the present formalism is the particular relation between the interaction vertex and \textquoteleft diamagnetic\textquoteright\  contributions to currents, such as $\gammak^{ij}=\partial_{k_i}(k_j\ekv)$ and $\gamma_{T,k}^{ij}= \frac{\partial}{\partial k_i} [k_j (\ekv)^2 ]$.
We finally show that these identities indicate that the total Hamiltonian, $H$, including the electric and thermal vector potentials are of the minimal form (to the second order in thermal vector potential)\cite{GT15minimal},
\begin{align}
 H&=\frac{\hbar^2}{2m}\sumkv\lt(\kv-e\Av-\epsilon_{k-eA}\ATv\rt)^2 c^\dagger_\kv c_\kv.
\label{minimalH}
\end{align} 
In fact, as is easily checked, $-\frac{\delta{H}}{\delta A_i}=\frac{e\hbar}{m}(k_i-eA_i-\gammak^{ij}A_{T,j})$ to the linear order of vector potentials ($\frac{\delta{H}}{\delta A_i}$ denotes a functional derivative), and thus the formal definition of current, $ \jv\equiv-\frac{\delta{H}}{\delta A_i}$, agrees with Eq. (\ref{jEresult}).
As for the energy current, it is formally defined by 
$j_{{\cal E},i} \equiv -\frac{1}{2} 
 \lt(\dot{c}^\dagger \frac{\delta H}{\delta (\nabla_i \cdag)}+ \frac{\delta H}{\delta (\nabla_i c)} \dot{c} \rt)$.
By use of the Heisenberg equation of motion for $\dot{c}_\kv$ and $\dot{c}^\dagger_\kv$,
we see that this formal definition agrees with Eq. (\ref{jEresult}).

In the electromagnetic case, the minimal form is imposed by a U(1) gauge invariance.
For the thermal vector potential, in contrast, there is no gauge invariance in the strict sense since the energy conservation arises from a global translational invariance with respect to time. \cite{GT15magnetic}
Still, our analysis indicates that the minimal form emerges.
This fact might be understood as due to a \textquoteleft gauge invariance\textquoteright\  as a result of the energy conservation law.
In fact, we have shown that the Luttinger's $\Psi$ and the present $\Av_T$ have the identical effect concerning steady state properties.
In other words, we may assign a part of thermal force to $\Psi$ and the rest to $\Av_T$, so that 
$\frac{\nabla T}{T}=\nabla\Psi+\dot{\Av}_T$.
Thus we have a \textquoteleft gauge invariance\textquoteright\  under a transformation $\Psi\ra\Psi-\dot{\chi}$ and $\Av_T\ra\Av_T+\nabla \chi$ ($\chi$ is a scalar function).
Such a gauge transformation is generally defined for a vector field coupling to a conserved current.

The thermal vector potential formalism applies to thermal torque straightforwardly
by replacing the electric charge in the derivation of  Ref. \cite{TE08} by energy.
It is easy to check that the thermal torque vanishes at $T=0$, as is physically required.

To conclude,
we have demonstrated that the calculation of transport coefficients are straightforwardly carried out based on a vector potential formalism. 
The formalism would apply straightforwardly to the case of multi-bands and to the interacting cases.

\begin{acknowledgements}
The author  thanks  H. Kohno, M. Ogata, H. Fukuyama, S.-K. Kim, A. Beekman, G. E. W. Bauer and N. Nagaosa 
for valuable comments.
This work was supported by a Grant-in-Aid for Scientific Research (C) (Grant No. 25400344), (A) (Grant No. 24244053) from the Japan Society for the Promotion of Science and  
Grant-in-Aid for Scientific Research on Innovative Areas (Grant No. 26103006) from The Ministry of Education, Culture, Sports, Science and Technology (MEXT), Japan.
\end{acknowledgements}


\begin{thebibliography}{20}
\bibitem[{GT1({\natexlab{a}})}]{GT15diamag}
\bibinfo{note}{Since the thermal force is conservative, the \textquoteleft
  diamagnetic\textquoteright current proportional to its vector potential does
  not have a diamagnetic effect. We still use the term \textquoteleft
  diagmagnetic\textquoteright to denote the current contribution proportional
  to the vector potential.}
\bibitem[{GT1({\natexlab{b}})}]{GT15AT}
\bibinfo{note}{We added an irrelavant constant $ \int_{-\infty}^{\infty} dt'
  \nabla\Psi(t') $ to the original form, $-\int_t^{\infty} dt'
  \nabla\Psi(t')$.}
\bibitem[{GT1({\natexlab{c}})}]{GT15curl}
\bibinfo{note}{The current determined from the continuity equation may have
  additional curl terms in general. Our result, however, indicates that such a
  term does not appear in the operator form of current in the present
  formalism, since the current turns out to be defined formally by the minimal
  coupling interaction as a sum of a paramagnetic and a \textquoteleft
  diamagnetic\textquoteright\ contributions. This fact is in contrast to the
  case of scalar potential, where the magnetization current written as a curl
  is essential as an equilibrium current contribution \cite{Smrcka77,Qin11}.}
\bibitem[{GT1({\natexlab{d}})}]{GT15minimal}
\bibinfo{note}{The minimal coupling of thermal vector potential is equivalent
  to the coupling between energy velocity and the vector potential in the
  Lagrangian representation.}
\bibitem[{GT1({\natexlab{e}})}]{GT15magnetic}
\bibinfo{note}{There is another difference between the thermal and the electric
  vector potentials: The former has no \textquoteleft magnetic\textquoteright\
  component and the \textquoteleft Lorentz force\textquoteright\ for energy
  current does not exist.}
\end{thebibliography}

\end{document}